\documentclass{ws-procs9x6-cpt19}
\begin{document}

\newcommand{\refeq}[1]{(\ref{#1})}
\def\etal {{\it et al.}}

\title{Exotic Spin-Dependent Interaction Searches \\ at Indiana University}

\author{I.\ Lee,$^1$ J.\ Shortino,$^1$ J.\ Biermen,$^1$ A.\ Din,$^1$ A.\ Grossman,$^1$ M.\ Gabel,$^1$ E.\ Guess,$^1$ C.-Y.\ Liu,$^1$ J.C.\ Long,$^1$ S.\ Reger,$^1$ A.\ Reid,$^1$ M.\ Severinov,$^1$ B.\ Short,$^1$ W.M.\ Snow,$^1$ E.\ Smith,$^1$ M.\ Zhang$^1$ and the ARIADNE Collaboration}

\address{$^1$Department of Physics, Indiana University,\\
Bloomington, IN 47405-7105, USA}

\begin{abstract}
The axion is a hypothesized particle appearing in various theories beyond the Standard Model. It is a light spin-0 boson initially postulated to solve the strong CP problem and is also a strong candidate for dark matter. If the axion or an axion-like particle exists, it would mediate a P-odd and T-odd spin-dependent interaction. We describe two experiments under development at Indiana University-Bloomington to search for such an interaction.
\end{abstract}

\bodymatter

\section{Introduction}
The Standard Model possesses many unexplained features. Why QCD does not violate CP is one of them. Peccei and Quinn in 1977 proposed that CP conservation in the strong interactions can be explained by axions.\cite{pecceiandquinn} Axions appear in many other beyond Standard Model theories, including string theory,\cite{stringtheory} and it is also considered as a strong candidate for dark matter.\cite{darkmatter}

Axion exchange generates a P-odd and T-odd spin-dependent potential energy which can be sought in laboratory experiments.\cite{IU1,IU2} Two new experiments with improved sensitivities are being developed. The PTB experiment exploits the world-renowned magnetically shielded room at Physikalisch-Technische Bundesanstalt (PTB) in Germany along with their high-developed SQUID magnetometry technology. The ARIADNE project is a collaboration among institutions in Korea, Canada, and the United States.

\section{Theory}
The axion would mediate a short-range monopole-dipole interaction with a potential of the form 

\begin{equation}
U(r)=\frac{\hbar^2 g_s g_p}{8\pi m_{f}}\left(\frac{1}{r\lambda_a}+\frac{1}{r^2}\right) e^{-(r/\lambda_a)}(\hat{\sigma}\cdot \hat{r}),
\end{equation}
where $g_s$ and $g_p$ are coupling constants, $m_{f}$ is fermion mass, $\hat{\sigma}$ is the Pauli spin matrix, $r$ is the distance between fermions, and $\lambda_a=h/m_a c$ is the axion Compton wavelength.\cite{moodyandwilczek} The interaction has Yukawa like potential so its strength drops quickly beyond the axion Compton wavelength. It affects the spin of dipole particles as magnetic field does, but since it is mediated by the axion, magnetic shielding has no effect on it.

\section{PTB Experiment}
The PTB experiment features a rotating disk with segments of alternating materials having similar magnetic properties but different nucleon densities, hence providing a time-varying axion field. A cell with hyperpolarized $^{3}$He and $^{129}$Xe will be placed near the mass, positioned to experience the axion field from only one material at a time. The axion field perpendicular to the longitudinal polarization axis of the samples causes the precession.

The resonant amplification of the signal gained by matching the frequency of the time-varying axion potential to the precession frequency of the hyperpolarized nuclei can greatly improve the sensitivity compared to previous experiments. The 8-layered magnetically shielded room at PTB\cite{ptb} allows a long spin relaxation time of the polarized gas samples on the order of 10,000 seconds to accumulate the resonant amplification effect. The transverse magnetization of precessing samples will be measured by PTB's highly sensitive superconducting quantum interference device (SQUID). The mixture of $^{3}$He and $^{129}$Xe can distinguish axion-mediated interaction from magnetic effects by treating one of the species as a comagnetometer. Nevertheless we suspect that magnetic impurities in the test masses will eventually pose a fundamental limitation on the sensitivity of this approach.

\section{Axion Resonant InterAction DetectioN Experiment (ARIADNE)}
The same experimental principles used in the PTB experiment are applied to the ARIADNE experiment. However, the ARIADNE experiment will be conducted at 4K in a liquid helium cryostat so that superconducting magnetic shielding can be employed to eliminate any test mass magnetic impurity systematics. The source mass is a rotating tungsten sprocket with 22 alternating segments. Three quartz blocks, each having a sample cell, bias coils, a SQUID loop, and niobium coating, will be placed next to the tungsten source mass. The blocks are thermally anchored to a 4K copper plate, turning its niobium coating into a superconducting shield against external magnetic backgrounds. The hyperpolarized $^3$He samples get cooled to 4K, increasing the density to improve the signal level. Such advantages of conducting the experiment at 4K add up to several orders of magnitude improvement in sensitivity. 

\section{ARIADNE at IU}
The $^3$He gas for ARIADNE is hyperpolarized by metastability exchange optical pumping (MEOP). Indiana University will provide the hyperpolarized $^3$He at 4K. A recycled liquid helium cryostat has been modified to house a Pyrex glass cell in which the entire process of polarizing and cooling $^3$He gas occurs. A complete MEOP system will be installed above the cryostat, and the polarized gas will diffuse down to the test cell into the 4K region. An NMR coil surrounding the test cell will be used to study the behavior of the polarized $^3$He samples at 4K. 

An RF shielded room is under construction at IU on a vibrationally isolated floor in the subbasement of the physics building. The floor is covered by RF shielding ferrite tiles and fine copper mesh. A Faraday cage with copper mesh will surround the experimental apparatus. This could be a possible experimental site for the ARIADNE project.

\section*{Acknowledgments}
We acknowledge support from the U.S. National Science Foundation, award numbers NSF PHY-1509176, NSF PHY-1509805, NSF PHY-1806395, NSF PHY-1806671, NSF PHY-1806757.

\end{document}